\documentclass[numberedappendix]{emulateapj} 

\usepackage{natbib}
\usepackage{hyperref}
\hypersetup{colorlinks,citecolor=Blue,linkcolor=Red,urlcolor=Blue}
\usepackage{amsmath}
\usepackage{empheq}
\usepackage{mathrsfs}
\usepackage[usenames,dvipsnames]{color}

\allowdisplaybreaks 

\begin{document}
 
\title{An Analytic Criterion for Turbulent Disruption of Planetary Resonances}  
\author{Konstantin Batygin$^{1}$ \& Fred C. Adams$^{2,3}$} 

\affil{$^1$Division of Geological and Planetary Sciences, California Institute of Technology, Pasadena, CA 91125} 
\affil{$^2$Department of Physics, University of Michigan, Ann Arbor, MI 48109} 
\affil{$^3$Department of Astronomy, University of Michigan, Ann Arbor, MI 48109} 

\email{kbatygin@gps.caltech.edu}
 
\newcommand{\Ham}{\mathcal{H}}
\newcommand{\G}{\mathcal{G}}
\newcommand{\D}{\mathcal{D}}
\newcommand{\taumig}{\tau_{\rm{mig}}}
\newcommand{\aave}{\langle a \rangle}
\newcommand{\appropto}{\mathrel{\vcenter{\offinterlineskip\halign{\hfil$##$\cr\propto\cr\noalign{\kern2pt}\sim\cr\noalign{\kern-2pt}}}}}

\begin{abstract} 
Mean motion commensurabilities in multi-planet systems are an expected
outcome of protoplanetary disk-driven migration, and their relative
dearth in the observational data presents an important challenge to
current models of planet formation and dynamical evolution. One natural
mechanism that can lead to the dissolution of commensurabilities is
stochastic orbital forcing, induced by turbulent density fluctuations
within the nebula. While this process is qualitatively promising, the
conditions under which mean motion resonances can be broken are not
well understood. In this work, we derive a simple analytic criterion that
elucidates the relationship among the physical parameters of the
system, and find the conditions necessary to drive planets out of resonance. Subsequently, we confirm our findings with
numerical integrations carried out in the perturbative regime, as well
as direct $N$-body simulations. Our calculations suggest that turbulent
resonance disruption depends most sensitively on the planet-star mass
ratio. Specifically, for a disk with properties comparable to the
early solar nebula with $\alpha=10^{-2}$, only planet pairs with cumulative mass ratios smaller than $(m_1+m_2)/M\lesssim10^{-5}\sim3M_{\oplus}/M_{\odot}$ are
susceptible to breaking resonance at semi-major axis of order
$a\sim0.1\,$AU. Although turbulence can sometimes compromise resonant
pairs, an additional mechanism (such as suppression of resonance
capture probability through disk eccentricity) is required to
adequately explain the largely non-resonant orbital architectures of
extrasolar planetary systems.
\end{abstract} 

\maketitle

\section{Introduction} \label{sect1}

Despite remarkable advances in the observational characterization of
extrasolar planetary systems that have occurred over the last two
decades, planet formation remains imperfectly understood. With the
advent of data from large-scale radial velocity and photometric
surveys \citep{Howard,Petigura,Batalha}, the origins of a newly
identified census of close-in Super-Earths (planets with orbital
periods that span days to months, and masses between those of the
Earth and Neptune) have emerged as an issue of particular
interest. Although analogs of such short-period objects are absent
from our solar system, statistical analyses have demonstrated that
Super-Earth type planets are extremely common within the Galaxy, and
likely represent the dominant outcome of planet formation \citep{Fressin2013,Forman-Mackey,Mulders2015}.

An elusive, yet fundamentally important aspect of the Super-Earth
conglomeration narrative is the role played by orbital transport.  
A key question is whether these planets experience accretion
\textit{in-situ} \citep{MurrayHansen,ChiangLaughlin,Lee2015,Lee2016},
or if they migrate to their close-in orbits having formed at large
orbital radii, as a consequence of disk-planet interactions
\citep{GoldreichTremaine1980,Tanaka2002,Crida2008,KleyNelson2012}.
Although this question remains a subject of active research, a number
recent studies \citep{Schlichting,Ogihara2015insitu,DAngeloBodenheimer2016} have pointed
to a finite extent of migration as an apparent requirement for
successful formation of Super-Earths. Moreover, structural models 
\citep{Rogers2015} show that the majority of Super-Earths have
substantial gaseous envelopes, implying that they formed in gas-rich
environments, where they could have actively exchanged angular
momentum with their surrounding nebulae.

\begin{figure}
\includegraphics[width=\columnwidth]{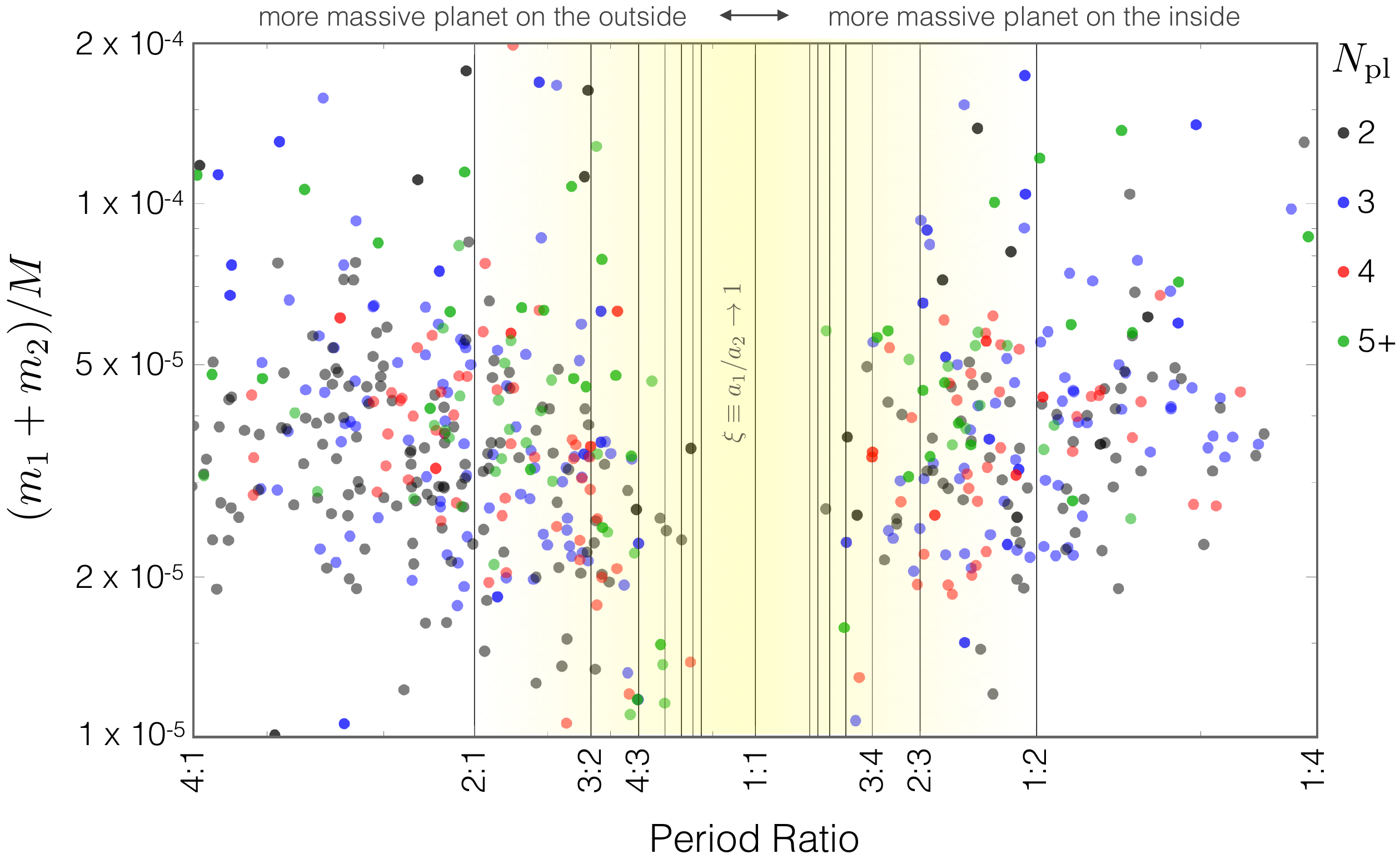}
\caption{Observed orbital distribution of Super-Earths. The ratio of  
orbital periods of confirmed planets is plotted against their
cumulative planet-star mass ratio. The period of the more massive
planet is adopted as a unit, such that systems that fall on the
left-hand-side of the 1:1 line have the more massive planet on the
outside, while the converse is true for systems that fall on the
right-hand-side of the 1:1 line. In systems where no direct
measurements of the mass (or $m\,\sin(i)$) are available, the mass is
inferred using the \citet{WeissMarcy2014} mass-radius
relationship. Such systems are shown with transparent points. The
planetary multiplicity, $N_{\rm{pl}}$, is color-coded in the following
way: systems with 2, 3, and 4 planets are shown with black, blue and
red points respectively. Systems with 5 or more planets are depicted
with green points. In planetary systems with more than two planets,
only period ratios of neighboring planets are considered. Vertical
lines denote the locations of first-order mean motion resonances. The
overall sample clearly shows little preference for orbital
commensurabilities.}
\label{data}
\end{figure}

Establishment of mean motion resonances in multi-planet systems has
long been recognized as a signpost of the planetary migration
paradigm. Specifically, the notion that slow, convergent evolution of
orbits towards one another produces planetary pairs with orbital
periods whose ratio can be expressed as a fraction of (typically
consecutive) integers, dates back more than half a century
\citep{Goldreich1965,Allan1969,Allan1970,Sinclair1970,Sinclair1972}.
While distinct examples of resonant planetary systems exist within the
known aggregate of planets\footnote{Archetypical examples of short-period resonant systems include GJ\,876 \citep{2010ApJ...719..890R}, Kepler-36 \citep{Decketal}, Kepler-79 \citep{2014ApJ...785...15J}, and Kepler-227 \citep{2016Natur.533..509M}.}, the
overall orbital distribution shows little preference for mean motion
commensurabilities (Figure \ref{data}). Therefore, taken at face
value, the paradigm of orbital migration predicts consequences for the
dynamical architectures of Super-Earths that are in conflict with the
majority of observations \citep{Fabrycky2014}. Accordingly,
\textit{the fact that mean motion commensurabilities are neither
  common nor entirely absent in the observational census of extrasolar
  planets presents an important challenge to the present understanding
  of planet formation theory}.

Prior to the detection of thousands of planetary candidates by the
\textit{Kepler} spacecraft, the expectations of largely resonant
architectures of close-in planets were firmly established by global
hydrodynamic, as well as $N$-body simulations
\citep{leepeale,quill,TerquemPap2007,CresswellNelson2008}. An important
distinction was drawn by the work of \cite{alb}, who pointed out that
resonances can be destabilized by random density fluctuations produced
by turbulence within the protoplanetary disk. Follow-up studies
demonstrated that a rich variety of outcomes can be attained as a
consequence of stochastic forcing within the disk
\citep{lecoanet,ketchum,lyra}, and that in specific cases, turbulence
can be conducive to the reproduction of dynamical architecture
\citep{rein,2010A&A...510A...4R}.

While the prediction of the infrequency of resonant systems made by
\cite{alb} was confirmed by the \textit{Kepler} dataset, recent work
has shown that turbulent forcing is not the only mechanism through
which resonances can be disrupted. Specifically, the work of
\cite{GoldShicht2014} proposed that a particular relationship between
the rates of eccentricity damping and semi-major axis decay can render
resonances metastable, while \cite{Batygin2015} showed that
probability of resonance capture can be dramatically reduced in
slightly non-axisymmetric disks. In light of the ambiguity associated
with a multitude of theoretical models that seemingly accomplish the
same thing, it is of great interest to inquire which, if any, of the
proposed mechanisms plays the leading role in sculpting the
predominantly non-resonant architectures of known exoplanetary
systems.

Within the context of the aforementioned models of resonant
metastability and capture suppression, the necessary conditions for
passage through commensurability are relatively clear. Resonant
metastability requires the outer planet to be much more massive than
the inner planet \citep{DeckBatygin}, while the capture suppression
mechanism requires disk eccentricities on the order of a few percent
to operate \citep{Batygin2015}. In contrast, the complex interplay
between planet-planet interactions, turbulent forcing, and dissipative
migration remains poorly quantified, making the turbulent disruption
mechanism difficult to decisively confirm or refute (see e.g., \citealt{ketchum}). As a result, a key goal of this work is to
identify the regime of parameter space for which the stochastic
dissolution of mean motion resonances can successfully operate. In
doing so, we aim to gain insight into the evolutionary stages of young
planetary systems during which disk turbulence can prevent the
formation of resonant pairs of planets.

The paper is organized as follows. In Section \ref{sect2}, we present
the details of our model. In Section \ref{sect3}, we employ methods
from stochastic calculus to derive an analytic criterion for
turbulent disruption of mean motion resonances. In Section
\ref{sect4}, we confirm our results with both perturbative numerical
integrations and an ensemble of full $N$-body simulations. The paper
concludes in Section \ref{sect5} with a summary of our results and a
discussion of their implications.

\section{Analytic Model} \label{sect2}

\begin{figure}
\includegraphics[width=\columnwidth]{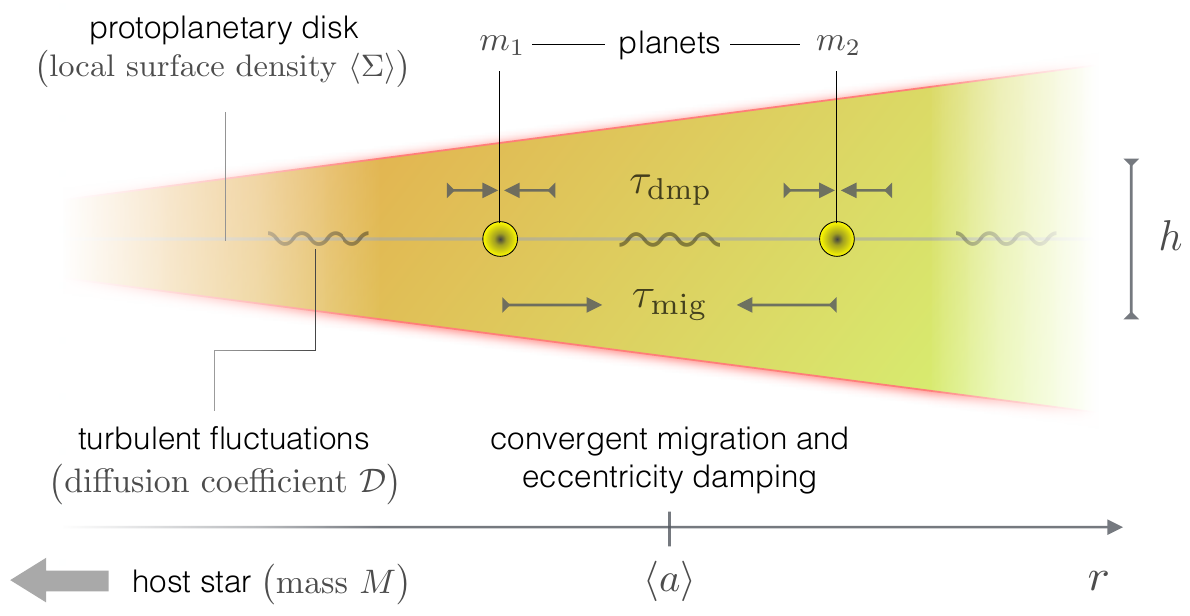}
\caption{Geometric setup of the dynamical model. Two planets with  
masses $m_1$ and $m_2$ are assumed to orbit a star of mass $M$ at an
approximate radial distance of $r=\langle a \rangle$. The bodies are
immersed in a gaseous nebula with scale-height $h$ and a nominal
local surface density $\langle \Sigma \rangle$. Tidal interactions
between the planets and the disk act to damp the planetary
eccentricities on a timescale $\tau_{\rm{dmp}}$, while facilitating
orbital convergence on a timescale $\tau_{\rm{mig}}$. Simultaneously,
turbulent density fluctuations within the nebula generate stochastic
perturbations to the planetary orbits, where the fluctuations are 
described by a diffusion coefficient $\mathcal{D}$.}
\label{setup}
\end{figure}

The model we aim to construct effectively comprises three ingredients:
(1) first-order ($k:k-1$) resonant planet-planet interactions, (2) orbital migration and
damping, as well as (3) stochastic turbulent forcing. In this section,
we outline our treatment of each of these processes. A cartoon
depicting the geometric setup of the problem is shown in Figure
\ref{setup}. Throughout much of the manuscript, we make the
so-called ``compact" approximation, where we assume that the
semi-major axis ratio $\xi \equiv a_1/a_2\rightarrow 1$. While
formally limiting, the agreement between results produced under this
approximation and those obtained within $N$-body integrations is
well-known to be satisfactory, particularly for $k\geqslant3$ (see, 
e.g., \citealt{Deck2013,DeckBatygin}), where the integer $k$ 
specifies the resonance \citep{md99,Morby}. 

Being made up of analytic components, the model constructed here
cannot possibly capture all of the intricate details of the dynamical
evolution that planets are subjected to, within protoplanetary disks.
By sacrificing precision on a detailed level, however, we hope to
construct an approximate description of the relevant physical
processes that will illuminate underlying relationships. These
findings can then be used to constrain the overall regime over which
turbulent fluctuations can effect the dynamical evolution of nascent
planetary systems.

\subsection{Planet-Planet Interactions}

In the late twentieth century, it was recognized that a perturbative
Hamiltonian that represents the motion of a massive pair of planets
residing on eccentric orbits, in the vicinity of a mean-motion
commensurability, can be cast into integrable form
\citep{SessinFerraz-Mello1984,Wisdom1986,Henrard1986}. More recently,
this formalism has been used to provide a geometric representation of
resonant dynamics \citep{BatMorby2013b}, study the onset of chaos
\citep{Deck2013}, generalize the theory of resonant capture
\citep{Batygin2015}, as well as to elucidate overstable librations
\citep{DeckBatygin}. A key advantage of this treatment is that it translates the full, unrestricted three-body problem into the same mathematical form as that employed for the well-studied circular restricted problem \citep{quill}. Here, we make use of this framework once again. Because detailed derivations of the aforementioned resonant normal
form are spelled out in the papers quoted above, we will not reproduce
it here, and instead restrict ourselves to employing the results. 

The Hamiltonian that describes planet-planet interactions in the
vicinity of a $k:k-1$ mean motion resonance can be written as follows: 
\begin{align}
\mathcal{H} = 3\,\big(\varepsilon+1\big) \bigg( \frac{x^2+y^2}{2} \bigg) - 
\bigg( \frac{x^2+y^2}{2} \bigg)^2 - 2\,x,
\label{Hammy}
\end{align}
where the variables $(\varepsilon,x,y)$ are defined below. 
A Hamiltonian of this form is typically referred to as the second
fundamental model for resonance
\citep{HenrardLemaitre1983,BordersGoldreich1984}, and behaves as a
forced harmonic oscillator at negative values of the proximity
parameter, $\varepsilon$, while possessing a pendulum-like phase-space
structure at large positive values of $\varepsilon$. This integrable
model approximates the real $N$-body dynamics at low eccentricities and
inclinations, and formally assumes that the orbits do not cross, 
although this latter assumption is routinely violated without much
practical consequence (see, e.g.,
\citealt{Peale1976,Malhotra1993,Deck2013}).  In the well-studied case
of the restricted circular three-body problem, the canonical variables
$(x,y)$ are connected to the test particle's eccentricity and the
resonant angle, while $\varepsilon$ is a measure of how close the
orbits are to exact resonance. Within the framework of the full
planetary resonance problem (where neither mass nor eccentricity of
either secondary body is assumed to be null), the variables take on
slightly more complex physical meanings.

In order to convert between Keplerian orbital elements and the
dimensionless canonical variables used here, we first define a
generalized composite eccentricity  
\begin{align}
&\sigma= \sqrt{e_1^2 + e_2^2- 2 e_1 e_2 \cos(\Delta \varpi)},
\end{align}
where subscripts $1$ and $2$ refer to the inner and outer planets
respectively, $e$ is eccentricity, and $\varpi$ is the longitude of
periastron. Additionally, we define units of action and time 
according to 
\begin{align}
&[A]= \frac{1}{2} \bigg( \frac{15}{4} 
\frac{k\,M}{m_1+m_2} \bigg)^{2/3} \nonumber \\
&[T] = \frac{1}{n}\bigg(\frac{5}{\sqrt{6}\,k^2} 
\frac{M}{m_1+m_2}\bigg)^{2/3},
\label{ATunits}
\end{align}
where $m$ is planetary mass, $M$ is stellar mass, and $n=\sqrt{\G\,M/a^3}$ 
is the mean motion. Then, in the compact limit, the variables in the 
Hamiltonian become \citep{DeckBatygin}:  
\begin{align}
x &= \sigma\, \bigg( \frac{15}{4} \frac{k\,M}{m_1+m_2} \bigg)^{1/3} 
\cos\big(k\lambda_2-(k-1)\lambda_1-\tilde{\omega}\big) \nonumber \\  
y &= \sigma\, \bigg( \frac{15}{4} \frac{k\,M}{m_1+m_2} \bigg)^{1/3} 
\sin\big(k\lambda_2-(k-1)\lambda_1-\tilde{\omega}\big) \nonumber \\
\varepsilon &= \frac{1}{3} \bigg( \frac{15}{4} \frac{k\,M}{m_1+m_2} 
\bigg)^{2/3} \bigg(\sigma^2 - \frac{\Delta \xi}{k} \bigg),
\label{AAcoords}
\end{align}
where $\xi=a_1/a_2$, and the quantity 
\begin{equation} 
\tilde{\omega} \equiv \arctan\left[
{e_2\sin\,\varpi_2-e_1\sin\,\varpi_1 \over e_1\cos\,\varpi_1-e_2\cos\,\varpi_2} 
\right] 
\end{equation} 
represents a generalized longitude of perihelion.

The specification of resonant dynamics is now complete. While
application of Hamilton's equations to equation (\ref{Hammy}) only
yields the evolution of $\sigma$ and the corresponding resonant angle,
the behavior of the individual eccentricities and apsidal lines can be
obtained from the conserved\footnote{When the system is subjected to
  slow evolution of the proximity parameter $\varepsilon$, $\rho$ is
  no longer a strictly conserved quantity. Instead, $\rho$ becomes an
  adiabatic invariant that is nearly constant, except when the system
  encounters a homoclinic curve \citep{BatMorby2013b}.}
quantity $\rho=m_1\,e_1^2+m_2\,e_2^2+m_1\,m_2\,e_1\,e_2\,\cos(\Delta
\varpi)$. In addition, we note that the definitions of the variables 
(\ref{AAcoords}) are independent of the individual planetary masses
$m_1, m_2$, and depend only on the cumulative planet-star mass ratio
$(m_1+m_2)/M$. This apparent simplification is a consequence of 
taking the limit $\xi \equiv a_1/a_2\rightarrow 1$, and is qualitatively
equivalent to the {\"O}pik approximation \citep{Opik1976}.

\subsection{Planet-Disk Interactions}

Dating back to early results on ring-satellite interactions
\citep{GoldreichTremaineRing}, it has been evident that planets can
exchange orbital energy and angular momentum with their natal
disks. For planets that are not sufficiently massive to open gaps
within their nebulae, this exchange occurs through local excitation 
of spiral density waves (i.e., the so-called ``type-I" regime), and
proceeds on the characteristic timescale: 
\begin{align}
&\tau_{\rm{wave}}= \frac{1}{n}\bigg(\frac{M}{m} \bigg) 
\bigg(\frac{M}{\Sigma\,a^2} \bigg) \bigg(\frac{h}{r}\bigg)^4,
\end{align}
where $\Sigma$ is the local surface density, and $h/r$ is the aspect
ratio of the disk. For an isothermal equation of state and a surface
density profile that scales inversely with the orbital radius 
\citep{Mestel63}, the corresponding rates of eccentricity and semi-major
axis decay are given by \citep{Tanaka2002,Tanaka2004}: 
\begin{align}
&\frac{1}{a}\frac{d\,a}{d\,t} \equiv - \frac{1}{\tau_{\rm{mig}}} 
\simeq - \frac{4\,f}{\tau_{\rm{wave}}} \bigg(\frac{h}{r}\bigg)^{2} \nonumber \\
&\frac{1}{e}\frac{d\,e}{d\,t}  \equiv - \frac{1}{\tau_{\rm{dmp}}} \simeq - 
\frac{3}{4} \frac{1}{\tau_{\rm{wave}}}.
\label{migdamp}
\end{align}
A different, routinely employed approach to modeling disk-driven
semi-major axis evolution is to assume that it occurs on a timescale
that exceeds the eccentricity decay time by a numerical factor
$\mathcal{K}$. To this end, we note that the value of $\mathcal{K}\sim10^2$ 
adopted by many previous authors \citep{leepeale,ketchum} is in rough 
agreement with equation (\ref{migdamp}) which yields 
$\mathcal{K}\sim(h/r)^{-2}$.

While eccentricity damping observed in numerical simulations 
(e.g., \citealt{CresswellNelson2008}) is well matched by equation
(\ref{migdamp}), state-of-the-art disk models show that both the rate
and direction of semi-major axis evolution can be significantly
affected by entropy gradients within the nebula \citep{Bitsch,Paardekooper}.
Although such corrections alter the migration histories on a detailed
level, convergent migration followed by resonant locking remains an
expected result in laminar disks \citep{ColemanNelson}.  For
simplicity, in this work, we account for this complication by
introducing an adjustable parameter $f$ into equation (\ref{migdamp}).

In addition to acting as sources of dissipation, protoplanetary disks
can also drive stochastic evolution. In particular, density
fluctuations within a turbulent disk generate a random gravitational
field, which in turn perturbs the embedded planets \citep{alb}.  Such
perturbations translate to effectively diffusive evolution of the
eccentricity and semi-major axis \citep{lsa,nelson}.  In the ideal
limit of MRI-driven turbulence, the corresponding eccentricity and
semi-major axis diffusion coefficients can be constructed from
analytic arguments (e.g., see \citealt{johnson,alb,OkuzumiOrmel2013}) 
to obtain the expressions 
\begin{align}
\D_{\xi} = \frac{\D_{a}}{a^2} \sim 2\,\D_{e} \sim \frac{\alpha}{2} 
\left( \frac{\Sigma\,a^2}{M} \right)^2 n,
\label{diffcoeff}
\end{align}
where $\alpha$ is the Shakura-Sunayev viscosity parameter 
\citep{ShakuraSunayev1973}. Although non-ideal effects can modify the
above expressions on the quantitative level \citep{okuzumi}, for the
purposes of our simple model we neglect these explicit corrections. We
note, however, that such details can be trivially incorporated into
the final answer by adjusting the value of $\alpha$ accordingly.

\section{Criterion for Resonance Disruption} \label{sect3}

With all components of the model specified, we now evaluate the
stability of mean motion resonances against stochastic
perturbations. In order to obtain a rough estimate of the interplay
between turbulent forcing, orbital damping, and resonant coupling, we
can evaluate the diffusive progress in semi-major axis and
eccentricity against the width of the resonance. Specifically, the
quantities, whose properties we wish to examine are $\chi \equiv
n_2/n_1 - k/(k-1)$ and $x$. Keep in mind that this latter quantity 
is directly proportional to the generalized eccentricity $\sigma$ 
(see equation [\ref{AAcoords}]).

\subsection{Diffusion of Semi-major Axes} 

In the compact limit $a_1\approx a_2$, the time evolution of the 
parameter $\chi$ can be written in the approximate form 
\begin{align}
\frac{d\chi}{dt}\simeq \frac{3}{2\,\aave}
\bigg(\frac{d a_1}{dt} - \frac{d a_2}{dt} \bigg),
\end{align}
where $\aave$ is a representative average semi-major axis. For the
purposes of our simple model, we treat $a_1$ and $a_2$ as uncorrelated
Gaussian random variables with diffusion coefficients $\D_{a}$; we note however, that in reality significant correlations may exist
between these quantities and such correlations could potentially alter the nature of the random walk \citep{rein}. Additionally, for comparable-mass planets, we
may adopt $\taumig$ as a characteristic drift rate, replacing $m$ with
$m_1+m_2$ in equation (\ref{migdamp}). Note that this assumption leads
to the maximum possible rate of orbital convergence.

With these constituents, we obtain a stochastic differential equation of the form
\begin{align}
d\chi = \frac{3}{2} \sqrt{2\,\D_{\xi}} \, dw - \frac{3}{2} \frac{\chi}{\taumig}\,dt,
\label{stochasticchi}
\end{align}
where $w$ represents a Weiner process (i.e., a continuous-time random
walk; \citealt{vankampen}). The variable $\chi$ will thus take on a distribution of values
as its evolution proceeds.  Adopting the $t\rightarrow \infty$
standard deviation of the resulting distribution function as a
characteristic measure of progress in $\chi$, we have: 
\begin{align}
\delta\chi = \sqrt{\frac{3\,\D_{\xi}\,\taumig}{2}} = 
\frac{1}{4} \frac{h}{r}\sqrt{\frac{3\,\alpha\,\Sigma\,\aave^2}{f\,(m_1+m_2)}}.
\end{align}

The approximate extent of stochastic evolution that the system can
experience and still remain in resonance is given by the resonant
bandwidth, $\Delta\chi$. At its inception\footnote{A resonance can
  only be formally defined when a homoclinic curve (i.e., a separatrix)
  exists in phase-space. For a Hamiltonian of the form (\ref{Hammy}),
  a separatrix appears at $\varepsilon=0$, along with an unstable
  (hyperbolic) fixed point, that bifurcates into two fixed points (one
  stable and one unstable) at $\varepsilon>0.$}, the width of the
resonance \citep{Batygin2015} is given by 
\begin{align}
\Delta\chi \simeq 5 \bigg[ \frac{\sqrt{k}\,(m_1+m_2)}{M} \bigg]^{2/3}.
\label{widthchi}
\end{align}
Accordingly, a rough criterion for turbulent disruption of the resonance is
\begin{empheq}[box=\fbox]{align}
\frac{\delta\chi}{\Delta\chi}&\sim\frac{1}{20}\frac{h}{r}
\frac{M}{m_1+m_2} \sqrt{\frac{3\,\alpha}{f}}\nonumber \\
&\times \Bigg[ \, \frac{\Sigma\,\aave^2}{k\,M} 
\sqrt{\frac{\Sigma\,\aave^2}{m_1+m_2}} \, \Bigg]^{1/3} \gtrsim 1.
\label{criterionchi}
\end{empheq}
Keep in mind that $\delta\chi$ is a measure of the width of the 
distribution in the variable $\chi$ due to stochastic evolution, 
whereas $\Delta\chi$ is the change in $\chi$ necessary to 
compromise the resonance. 

\begin{figure}
\includegraphics[width=\columnwidth]{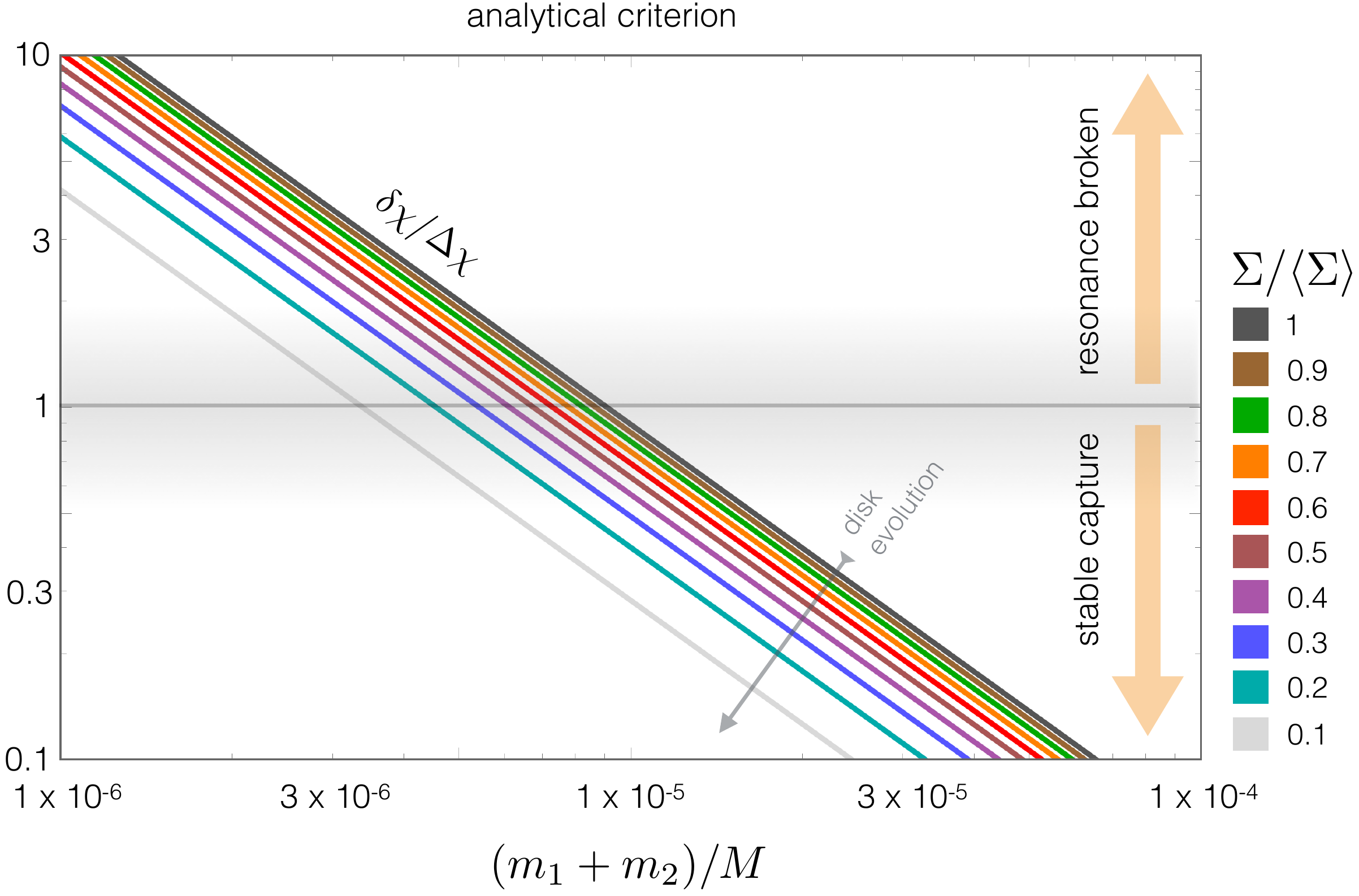}
\caption{Analytic criterion for resonance disruption. Expression
(\ref{criterionchi}) is shown as a function of the cumulative  
planet-star mass ratio, $(m_1+m_2)/M$. Resonances are stable against
stochastic perturbations in the region of the graph where
$\delta\chi/\Delta\chi\ll1$ and are unstable where
$\delta\chi/\Delta\chi\gg1$. Notably, $\delta\chi/\Delta\chi\sim1$
represents a transitional regime, where resonance capture may
successfully occur, but will generally not be permanent. In this
example, the disk viscosity parameter and the disk aspect ratio are
taken to be $\alpha=10^{-2}$ and $h/r=0.05$, respectively. The planets
are envisioned to reside at $\langle a \rangle=0.1\,$AU, in a nebula
with a nominal local surface density $\langle\Sigma\rangle$ = 17,000 
g cm$^{-2}$. A family of curves corresponding to lower values of the 
surface density are also shown, and color-coded accordingly. Finally,
the migration parameter and the resonance index are set to $f=1$ and
$k=3$, respectively. } 
\label{critfig}
\end{figure}

The above expression for resonance disruption depends sensitively on
the planet-star mass ratio. This relationship is illustrated in Figure
\ref{critfig}, where the expression (\ref{criterionchi}) is
shown as a function of the quantity $(m_1+m_2)/M$, assuming system
properties $\alpha=10^{-2}$, $h/r=0.05$, $\langle a\rangle = 0.1\,$AU,
$f=1$, and $k=3$. The disk profile is taken to have the form
$\Sigma=\Sigma_0\,(r_0/r)$, with $\Sigma_0=1700\,\rm{g/cm}^{2}$ and
$r_0=1\,$AU, such that the local surface density at $\langle a\rangle$
is $\langle \Sigma \rangle = 17,000\,\rm{g/cm}^{2}$. Notice that the 
disruption criterion (\ref{criterionchi}) also depends on (the square
root of) the surface density of the disk.  A family of curves
corresponding to lower values of the surface density (i.e.,
$0.1,0.2,\dots0.9,1\,\times\,\langle \Sigma \rangle$) are also shown,
and color-coded accordingly.

While Figure \ref{critfig} effectively assumes a maximal rate of orbital convergence, we reiterate that hydrodynamical simulations suggest that both the speed and sense of type-I migration can have a wide range of possible values \citep{Paardekooper}. To this end, we note that setting $f=0$ in equation (\ref{criterionchi}) yields $\infty>1$, meaning that in the case of no net migration, an arbitrarily small turbulent viscosity is sufficient to eventually bring the resonant angles into circulation. Furthermore, a negative value of $f$, which corresponds to divergent migration, renders our criterion meaningless, since resonance capture cannot occur in this instance \citep{Peale1976}.

\subsection{Diffusion of Eccentricities} 

An essentially identical calculation can carried out for stochastic
evolution of $x$ (or $y$). To accomplish this, we assume that the
generalized eccentricity $\sigma$ diffuses with the coefficient
$\sqrt{2}\,\D_{e}$. Accounting for conversion factors between
conventional quantities and the dimensionless coordinates (given by
equation [\ref{ATunits}]), we obtain 
\begin{align}
\D_{x} \simeq \alpha\,\bigg(\frac{\Sigma\,\aave^2}{M} \bigg)^2 
\bigg(\frac{M}{\sqrt{k}\,(m_1+m_2)}\bigg)^{4/3}.
\end{align}
Similarly, the damping timescale takes the form 
\begin{align}
\tau_{x} \simeq \, \bigg(\frac{128}{225} 
\frac{k\, M}{m_1+m_2} \bigg)^{1/3} \frac{k\,M}{\Sigma\,\aave^2} \bigg(\frac{h}{r}\bigg)^4,
\end{align}
where, as before, we adopted the total planetary mass as an an
approximation for $m$ in the expression (\ref{migdamp}).

In direct analogy with equation (\ref{stochasticchi}), we obtain 
the stochastic equation for the time evolution of $x$, 
\begin{align}
dx = \sqrt{2\,\D_{x}} \, dw - \frac{x}{\tau_{x}}\,dt,
\end{align}
so that the distribution of $x$ is characterized by the standard 
deviation $\delta x = \sqrt{\D_x\,\tau_x}$. At the same time, we
take the half-width of the resonant separatrix to be given by 
$\Delta x=2$ (e.g., \citealt{BatMorby2013b,Decketal}). Combining 
these two results, we obtain a second criterion for resonance 
disruption, i.e., 
\begin{align}
\frac{\delta x}{\Delta x}&\sim \bigg(\frac{h}{r}\bigg)^2 \, 
\bigg(\frac{\sqrt{2}\,k}{15} \bigg)^{1/3} \sqrt{\alpha \, 
\frac{\Sigma\,\aave^2}{M}} \nonumber \\
&\times \bigg(\frac{M}{m_1+m_2} \bigg)^{5/6} \gtrsim 1 \,. 
\label{criterione}
\end{align}

\subsection{Semi-major Axis vs Eccentricity} 

In order to construct the simplest possible model that still captures
the dynamical evolution adequately, it is of interest to evaluate the
relative importance of stochastic evolution in the degrees of freedom
related to the semi-major axis and eccentricity. Expressions
(\ref{criterionchi}) and (\ref{criterione}) both represent conditions
under which resonant dynamics of a planetary pair will be short-lived,
even if capture occurs. To gauge which of the two criteria is more
stringent, we can examine the ratio 
\begin{align}
\frac{\delta x/\Delta x}{\delta\chi/\Delta\chi} \sim  
{5}{\sqrt{f}}\, \bigg( \frac{h}{r} \bigg) \, 
\bigg[ \frac{k^2\,(m_1+ m_2)}{M} \bigg]^{1/3} \ll 1.
\label{avse}
\end{align}
The fact that this expression evaluates to a number substantially
smaller than unity means that diffusion in semi-major axes (equation
[\ref{criterionchi}]) dominates over diffusion in eccentricities
(equation [\ref{criterione}]) as a mechanism for disruption of
mean-motion commensurabilities. Although the relative importance of
$\D_{a}$ compared to $\D_{e}$ is not obvious {\it a priori}, it likely
stems in large part from the fact that the orbital convergence
timescale generally exceeds the eccentricity damping timescales by a
large margin.

\begin{figure}
\includegraphics[width=1\columnwidth]{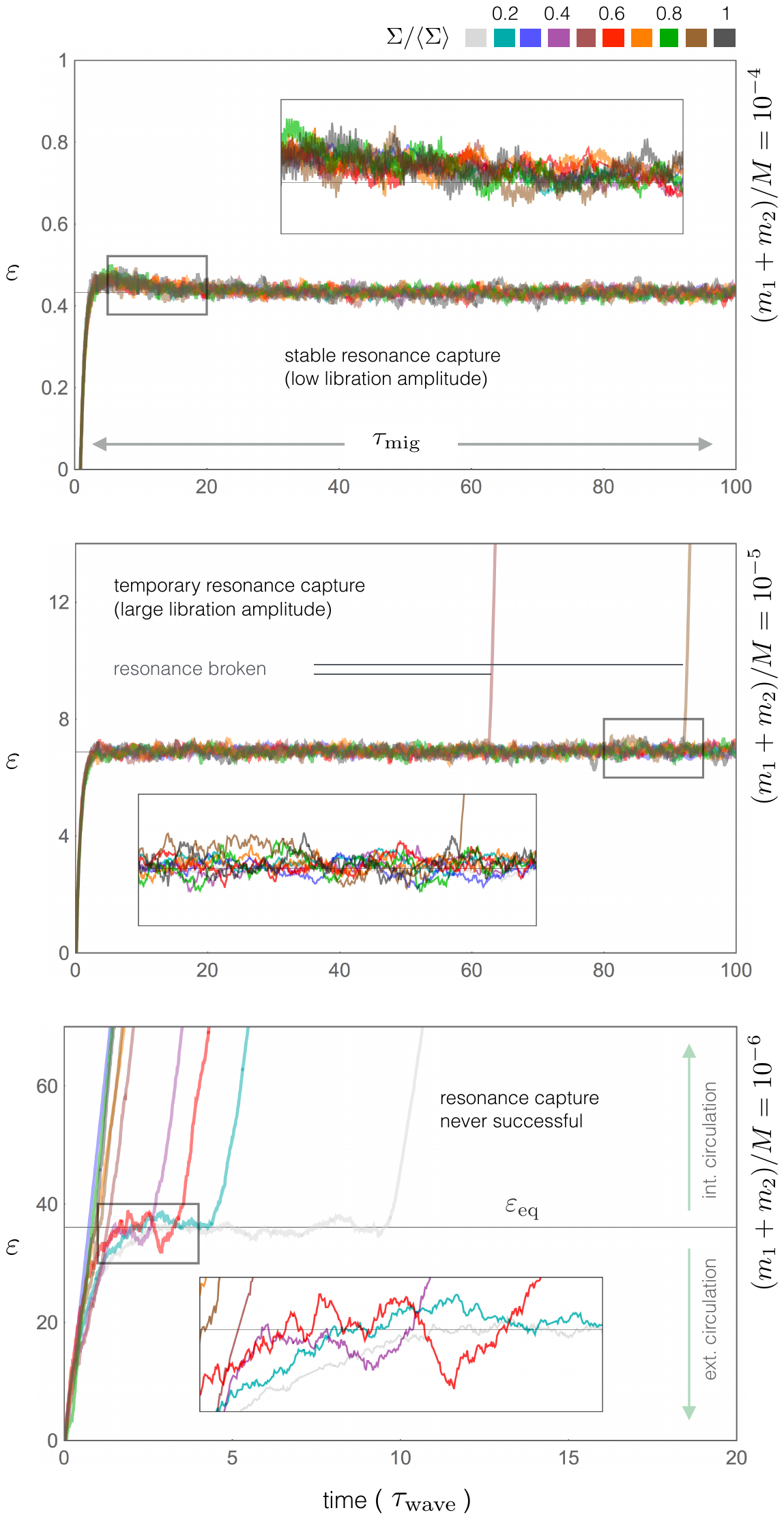}
\caption{Evolution of the resonance proximity parameter. The top,  
middle, and bottom panels of the Figure correspond to cumulative
planet-star mass ratios of $(m_1+m_2)/M=10^{-4}$, $10^{-5}$, and
$10^{-6}$ respectively. On each panel, ten simulations corresponding
to different local surface densities 
($0.1,0.2,\dots0.9,1\,\times\langle \Sigma \rangle$; color-coded
accordingly) are shown. In agreement with the analytic criterion
(equation \ref{criterionchi}), systems less massive than
$(m_1+m_2)/M\lesssim10^{-5}$ are susceptible to turbulent resonance
disruption, while systems with $(m_1+m_2)/M\gtrsim10^{-5}$ experience
stable resonance capture. Two out of ten simulations with
$(m_1+m_2)/M=10^{-5}$ show resonance breaking, implying that for the
adopted set of physical parameters, this mass ratio corresponds to
critical behavior. Importantly, systems of this type can emerge from
the protoplanetary disk with large resonant libration amplitudes.}
\label{delta}
\end{figure}

\section{Numerical Integrations} \label{sect4}

In order to derive a purely analytic criterion for turbulent
disruption of mean motion resonances, we were forced to make a series
of crude approximations in the previous section. To assess the
validity of these approximations, in this section we test the
criterion (\ref{criterionchi}) through numerical integrations.  We
first present a perturbative approach (Section \ref{perturb}) and then
carry out a series of full $N$-body simulations (Section \ref{nbody}).

\subsection{Perturbation Theory}
\label{perturb} 

The dynamical system considered here is described by three equations
of motion, corresponding to the variations in $x$, and $y$, and
$\varepsilon$. Although the resonant dynamics itself is governed by
Hamiltonian (\ref{Hammy}), to account for the stochastic and
dissipative evolution, we must augment Hamilton's equations with terms
that describe disk-driven evolution. As before, we adopt
$\tau_{\rm{dmp}}$ as the decay timescale for the generalized
eccentricity, $\sigma$, and take $\tau_{\rm{mig}}$ as the
characteristic orbital convergence time. The full equations of motion
are then given by:
\begin{align}
&\frac{dx}{dt} = -3\,y\,\big(1+\varepsilon \big)+y\,\big(x^2+y^2\big) - \frac{x}{\tau_{\rm{dmp}}/[T]}  \nonumber \\
&\frac{dy}{dt} = -2+3\,x\,\big(1+\varepsilon\big) - x\,\big(x^2 + y^2 \big) - \frac{y}{\tau_{\rm{dmp}}/[T]} \nonumber \\
&\frac{d\varepsilon}{dt} = \frac{2}{3}\bigg(\frac{[A]}{k\,\tau_{\rm{mig}}/[T]} - \frac{x^2+y^2}{\tau_{\rm{dmp}}/[T]} \bigg)+\mathcal{F}.
\label{EOM}
\end{align}

In the above expression, $\mathcal{F}$ represents a source of
stochastic perturbations. For computational convenience, we
implemented this noise term as a continuous sequence of analytic
pulses, which had the form $2\,\zeta \sin(\pi\,t/\Delta t )/\Delta t$, 
where $\zeta$ is a Gaussian random variable. The pulse time interval
was taken to be $\Delta t=0.1$, and the standard deviation of $\zeta$
was chosen such that the resulting diffusion coefficient
$\mathcal{D}_{\zeta} = \sigma_{\zeta}^2/\Delta t$ matched that given
by equation (\ref{diffcoeff}).

\begin{figure*}
\includegraphics[width=\textwidth]{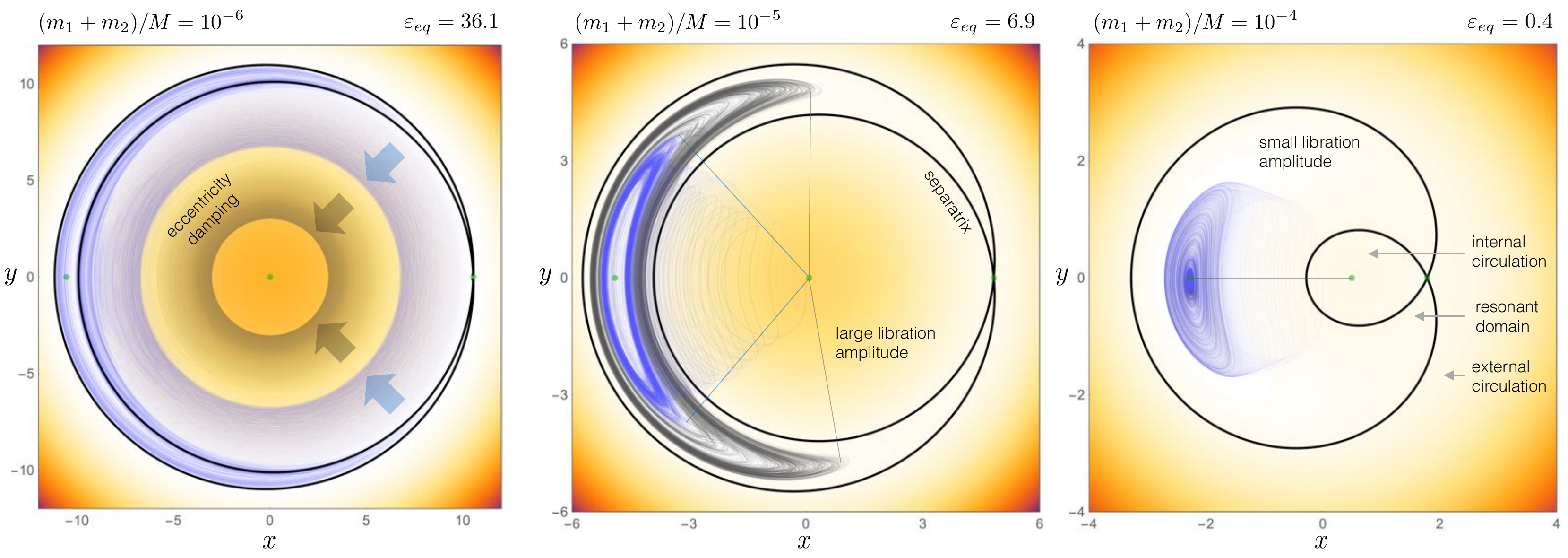}
\caption{Numerically determined phase-space evolution of the dynamical
  system. As in Figure \ref{delta}, the left, middle, and right
  panels correspond to cumulative planet-star mass ratios of
  $(m_1+m_2)/M=10^{-6}$, $10^{-5}$, and $10^{-4}$
  respectively. Simulation results for $\Sigma=\langle \Sigma \rangle$
  (black) and $\Sigma= 0.3\,\langle \Sigma \rangle$ (blue) are
  shown. In each plot, the solid black line depicts the separatrix of
  the Hamiltonian (\ref{Hammy}), evaluated at the equilibrium
  proximity parameter, $\varepsilon_{\rm{eq}}$, while the color scale
  denotes level curves of $\mathcal{H}$. In the left panel, the
  trajectories are initially advected to large actions, but eventually
  break out of resonance and begin decaying towards the fixed point at
  the center of the internal circulation region of the dynamical
  portrait. The middle panel shows a critical evolution where
  stochastic excursions of the trajectories are limited by dissipation
  to fill a substantial fraction of the resonant phase-space, without
  breaking out of resonance. Systems in this parameter range can
  emerge from the nebula with large resonant libration amplitudes,
  potentially leading to chaotic evolution. The right panel shows an
  evolutionary sequence where stochastic forcing plays an essentially
  negligible role, i.e., the proximity parameter equilibrates and the
  orbit collapses onto the resonant equilibrium under the action of
  dissipative effects.}
\label{pspacefig}
\end{figure*}

Note that here, we have opted to only implement stochastic
perturbations into the equation that governs the variation of
$\varepsilon$. Qualitatively, this is equivalent to only retaining
semi-major axis diffusion and neglecting eccentricity diffusion. To
this end, we have confirmed that including (appropriately scaled)
turbulent diffusion into equations of motion for $x$ and $y$ does not
alter the dynamical evolution in a meaningful way, in agreement with
the discussion surrounding equation (\ref{avse}).

Turbulent fluctuations aside, the equation of motion for the parameter
$\varepsilon$ indicates that there exists an equilibrium value of the
generalized eccentricity
$\sigma_{\rm{eq}}=\sqrt{\tau_{\rm{dmp}}/(2\,k\,\tau_{\rm{mig}})}$ that
corresponds to stable capture into resonance. Analogously, the
equilibrium value of $x_{\rm{eq}}=\sigma_{\rm{eq}} \sqrt{2[A]}$
parallels the strictly real fixed point of Hamiltonian (\ref{Hammy}).
As a result, if we neglect the small dissipative contributions and set
$dx/dt=0,dy/dt=0,x=x_{\rm{eq}},y=0$ in the first and second equations
in expression (\ref{EOM}), we find an equilibrium value of the
proximity parameter, $\varepsilon_{\rm{eq}}$, that coincides with
resonant locking. An ensuing crucial point is that if resonance is
broken, the system will attain values of $\varepsilon$ substantially
above the equilibrium value $\varepsilon_{\rm{eq}}$.

In order to maintain a close relationship with the results presented
in the preceding section, we retained the same physical parameters for
the simulations as those depicted in Figure \ref{critfig}. In
particular, we adopted $\alpha=10^{-2}$, $h/r=0.05$, $\langle a
\rangle=0.1\,$AU, $f=1$, and $k=3$. Additionally, we again chose a
surface density profile with $\Sigma_0=1700\,$g/cm$^{2}$ at
$r_0=1\,$AU, that scales inversely with the orbital radius, such that
the nominal surface density at $r=\langle a \rangle$ is $\langle
\Sigma \rangle=17,000\,$g/cm$^{2}$. We also performed a series of
simulations that span a lower range of surface densities
($0.1,0.2,\dots,0.9,1\,\times\,\langle \Sigma \rangle$). 
All of the integrations were carried out over a time span of
$\tau_{\rm{mig}}=100\,\tau_{\rm{wave}}$, with the system initialized
at zero eccentricity ($x_0=y_0=0$), on orbits exterior to exact
commensurability ($\epsilon_0=-1$).

We computed three sets of evolutionary sequences, corresponding to
planet-star mass ratios $(m_1+m_2)/M=10^{-6},10^{-5},$ and $10^{-4}$.
As can be deduced from Figure \ref{critfig}, the qualitative
expectations for the outcomes of these simulations (as dictated by
equation [\ref{criterionchi}]) are unequivocally clear. Resonances
should be long-term stable for $(m_1+m_2)/M=10^{-4}$ and long-term
unstable for $(m_1+m_2)/M=10^{-6}$. Meanwhile, temporary resonance
locking, followed by turbulent disruption of the commensurability
should occur for $(m_1+m_2)/M=10^{-5}$.

Figure \ref{delta} depicts numerically computed evolution of
$\varepsilon$ for the full range of local surface densities under
consideration (color-coded in the same way as in Figure \ref{critfig})
as a function of time. These numerical results are in excellent
agreement with our theoretical expectations from Section \ref{sect3}.
The proximity parameter always approaches its expected equilibrium
value $\varepsilon_{\rm{eq}}$ for large mass ratios
$(m_1+m_2)/M=10^{-4}$ (top panel), but never experiences long-term
capture for small mass ratios $(m_1+m_2)/M=10^{-6}$ (bottom
panel). Resonance locking does occur for the intermediate case
$(m_1+m_2)/M=10^{-5}$ (middle panel). However, two evolutionary
sequences corresponding to $\Sigma=0.7\,\langle \Sigma \rangle$ and
$\Sigma=0.9\,\langle \Sigma \rangle$ show the system breaking out of
resonance within a single orbital convergence time, $\tau_{\rm{mig}}$.
It is sensible to assume that other evolutionary sequences within this
set would also break away from resonance if integrations were
extended over a longer time period.

Figure \ref{pspacefig} shows the phase-space counterpart of the
evolution depicted in Figure \ref{delta}. Specifically, the $x$-$y$
projections of the system dynamics are shown for cases with surface
densities $\Sigma=0.3\,\langle \Sigma \rangle$ (blue) and
$\Sigma=\langle \Sigma \rangle$ (black), where the background depicts
the topology of the Hamiltonian (\ref{Hammy}). In each panel, the
black curve designates the separatrix of $\mathcal{H}$, given the
equilibrium value of the proximity parameter
$\varepsilon=\varepsilon_{\rm{eq}}$. The background color scale is a
measure of the value of $\mathcal{H}$. The thee equilibrium points of
the Hamiltonian are also shown, as transparent green dots.

As in Figure \ref{delta}, three representative ratios of
planet mass to stellar mass are shown.  In the right panel (for mass ratio
  $(m_1+m_2)/M=10^{-4}$), turbulent diffusion plays an essentially
negligible role and the system approaches a null libration amplitude 
under the effect of dissipation. In the middle panel (for mass ratio
  $(m_1+m_2)/M=10^{-5}$), resonant capture is shown, but the libration
amplitude attained by the orbit is large, particularly in the case of
$\Sigma=\langle \Sigma \rangle$. In the left panel (for mass ratio
  $(m_1+m_2)/M=10^{-6}$), the trajectory is initially advected to high
values of the action, but inevitably breaks out of resonance and
decays towards the fixed point at the center of the internal
circulation region of the portrait.

\subsection{$N$-body Simulations}
\label{nbody} 

In order to fully evaluate the approximations inherent to the
perturbative treatment of the dynamics employed thus far, and to
provide a conclusive test of the analytic criterion
(\ref{criterionchi}), we have carried out a series of direct $N$-body
simulations. The integrations utilized a Burlisch-Stoer
integration scheme (e.g., \citealt{Press1992}) that included the full
set of 18 phase space variables for the 3-body problem consisting of
two migrating planets orbiting a central star.  For the sake of
definiteness, the physical setup of the numerical experiments was
chosen to closely mirror the systems used in the above discussion.
Specifically, two equal-mass planets were placed on initially circular
orbits slightly outside of the 2:1 mean motion resonance, so that the
initial period ratio was $0.45$. The planets were then allowed to
evolve under the influence of mutual gravity, as well as disk-driven
convergent migration, orbital damping, and turbulent perturbations.

\begin{figure}
\includegraphics[width=\columnwidth]{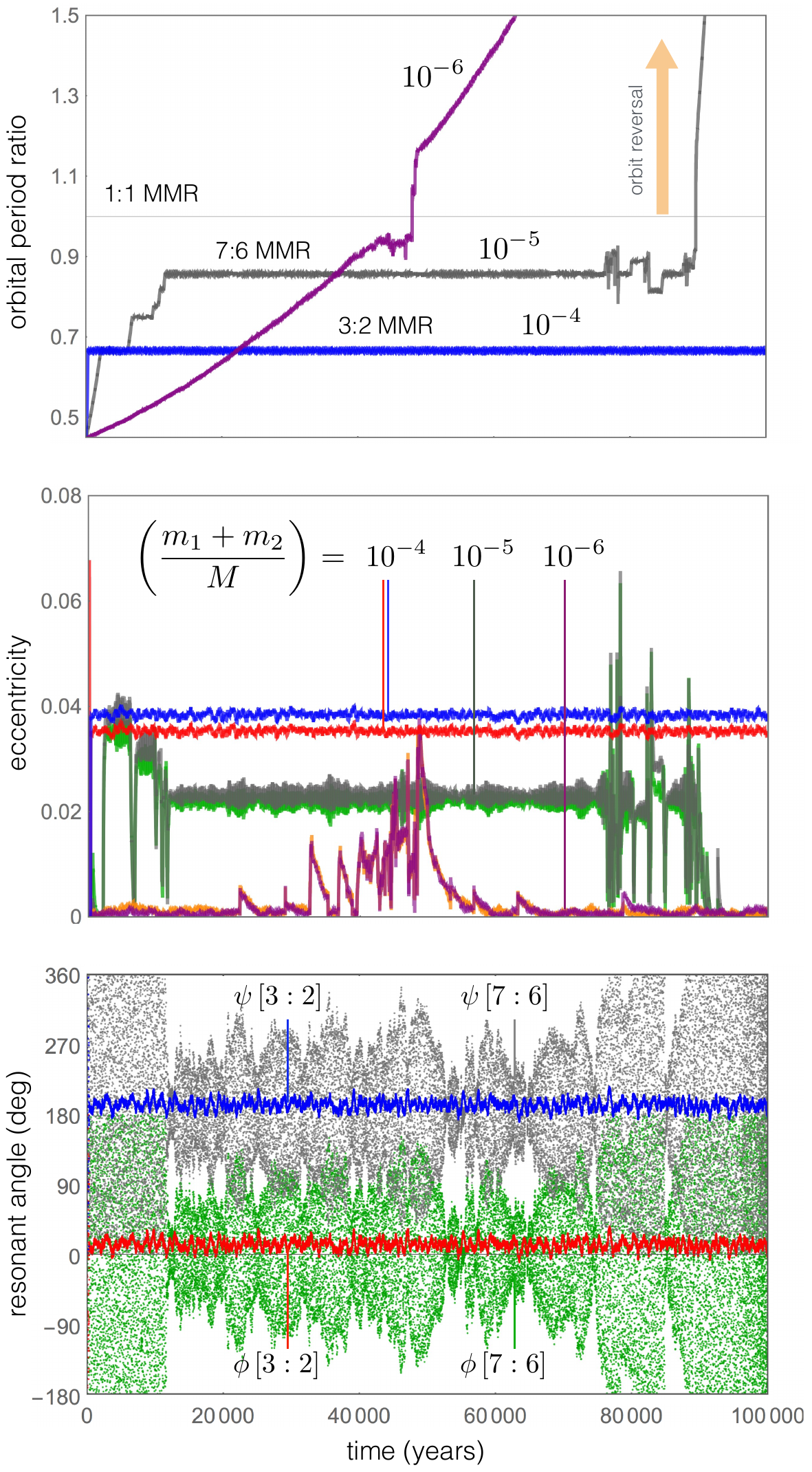}
\caption{Results of direct $N$-body simulations. This figure shows the  
time series of the orbital period ratio (top), eccentricities
(middle), and resonant angles (bottom) for a pair of planets subject
to convergent migration, eccentricity damping, and stochastic forcing. 
The disk is assumed to be comparable to the minimum mass solar nebula
and the planetary orbits lie at $a\sim0.1\,$AU. Three representative
sets of evolutionary tracks are shown with mass ratios 
$(m_1+m_2)/M=10^{-4},10^{-5},$ and $10^{-6}$. In the top panel, the
curves corresponding to planet-star mass ratios of
$(m_1+m_2)/M=10^{-4},10^{-5},$ and $10^{-6}$ are shown in blue, gray,
and purple respectively. In the middle panel, the eccentricities for
$(m_1+m_2)/M=10^{-4}$ are shown as blue (outer planet) and red (inner
planet) curves. Similarly, the gray and green as well as purple and
orange curved denote the eccentricities of outer and inner planets for
$(m_1+m_2)/M=10^{-5}$ and $10^{-6}$. The bottom panel shows resonant
arguments $\phi$[3:2] = $3\lambda_2 - 2 \lambda_1-\varpi_1$ (red) and
$\psi$[3:2] = $3\lambda_2 - 2 \lambda_1-\varpi_2$ (blue) corresponding
to the system with $(m_1+m_2)/M=10^{-4}$ as well as resonant arguments
$\phi$[7:6] = $7\lambda_2 - 6\lambda_1-\varpi_1$ (green) and
$\psi$[7:6] = $3\lambda_2 - 2 \lambda_1-\varpi_2$ (gray) corresponding
to the system with $(m_1+m_2)/M=10^{-5}$. In agreement with the
analytic criterion (\ref{criterionchi}), the system with
$(m_1+m_2)/M=10^{-4}$ exhibits stable capture into a 3:2 resonance,
while the system with $(m_1+m_2)/M=10^{-5}$ only becomes temporarily
trapped into a 7:6 commensurability before breaking out due to
turbulent perturbations.  Conversely, the system with
$(m_1+m_2)/M=10^{-6}$ never locks into resonance and eventually
suffers orbit reversal.}
\label{Nbodyfig}
\end{figure}

Following \citet{PapaloizouLarwood2000}, we incorporated the
orbital decay and eccentricity damping using accelerations 
of the form: 
\begin{align}
&\frac{d\vec{v}}{dt} = -\frac{\vec{v}}{\tau_{\rm{mig}}} - 
\frac{2\,\vec{r}}{\tau_{\rm{dmp}}} 
\frac{\left(\vec{v}\cdot{\vec{r}}\right)}{\left(\vec{r} \cdot \vec{r} \right)},
\label{PapLar}
\end{align}
where $\vec{v}$ and $\vec{r}$ denote the orbital velocity and radius
respectively.\footnote{Note that we have neglected disk-induced damping 
of the orbital inclination, because of the planar setup of the problem.}  
While both planets were subjected to eccentricity
damping, inward (convergent) migration was only experienced by the
outer planet. Simultaneously, for computational convenience, the
semi-major axis of the inner planet was re-normalized to $a_1=0.1\,$AU
at every time step\footnote{Qualitatively, this procedure is equivalent
to changing the unit of time at every time step \citep{DeckBatygin}.}.  
The characteristic timescales $\tau_{\rm{mig}}$ and $\tau_{\rm{dmp}}$
were kept constant, given by equation (\ref{migdamp}), adopting
identical physical parameters of the disk to those employed above. 
Finally, following previous treatments \citep{alb,rein,lecoanet}, 
turbulent fluctuations were introduced into the equations of motion
through random velocity kicks, whose amplitude was tuned such that the
properties of the diffusive evolution of an undamped isolated orbit
matched the coefficients from equation (\ref{diffcoeff}). For completeness, we have also included the leading order corrections due to general relativity \citep{genrel}.

As in the previous sub-section, we computed the orbital evolution of
three representative cases with mass ratios
$(m_1+m_2)/M=10^{-4},10^{-5},$ and $10^{-6}$ (corresponding to migration timescales of $\tau_{\rm{mig}}\simeq1.5\times10^3,10^4$, and $10^5$\, years respectively) over a time span of
$0.1\,$Myr. The numerical results are shown in Figure \ref{Nbodyfig},
and show excellent agreement with the analytic criterion from equation
(\ref{criterionchi}). In particular, the system with mass ratio 
$(m_1+m_2)/M=10^{-4}$ exhibits long-term stable capture into a 3:2
mean motion resonance, as exemplified by the ensuing low-amplitude
libration of the resonant angles $\phi$[3:2] = $3\lambda_2 - 2
\lambda_1-\varpi_1$ and $\psi$[3:2] = $3\lambda_2 - 2 \lambda_1-\varpi_2$,
shown in red and blue in the bottom panel of Figure \ref{Nbodyfig}. 
Correspondingly, both the period ratio (top panel) and the
eccentricities (middle panel) rapidly attain their resonant
equilibrium values, and remain essentially constant throughout the
simulation.

The case with mass ratio $(m_1+m_2)/M=10^{-5}$, for which equation
(\ref{criterionchi}) yields $\delta \chi / \Delta \chi \sim1$,
perfectly exemplifies the transitory regime. As shown in the top
panel of Figure \ref{Nbodyfig}, where this experiment is represented
in gray, the system exhibits temporary capture into the 3:2 as well
as the 4:3 commensurabilities, and subsequently locks into a
meta-stable 7:6 resonance at time $\sim15,000$ years. Although evolution
within this resonance is relatively long-lived, the bottom panel of
Figure \ref{Nbodyfig} shows that the corresponding resonant angles
$\phi$[7:6] = $7\lambda_2 - 6\lambda_1-\varpi_1$ (green) and
$\psi$[7:6] = $3\lambda_2 - 2 \lambda_1-\varpi_2$ (gray) maintain 
large amplitudes of libration, due to the nearly perfect balance
between orbital damping and turbulent excitation. As a result, the
system eventually breaks out of its resonant state. After a period of
chaotic scattering, the orbits switch their order, and the period
ratio increases.

Finally, the case with mass ratio $(m_1+m_2)/M=10^{-6}$ represents a
system that never experiences resonant locking. As the period ratio
evolves towards unity (purple curve in the top panel), encounters with
mean motion commensurabilities only manifest themselves as impulsive
excitations of the orbital eccentricities (purple/orange curves in the
middle panel) of the planets. As such, the planets eventually
experience a brief phase of close encounters, and subsequently
re-enter an essentially decoupled regime, after the orbits reverse. 

We note that because turbulence introduces a fundamentally stochastic
component into the equations of motion, each realization of the $N$-body
simulations is quantitatively unique. However, having carried out tens
of integrations for each set of parameters considered in Figure
\ref{Nbodyfig}, we have confirmed that the presented solutions are
indeed representative of the evolutionary outcomes. As a result, we 
conclude that the analytic expression (\ref{criterionchi}) represents
an adequate description of the requirement for resonance disruption, 
consistent with the numerical experiments.

\section{Conclusion} \label{sect5}

While resonant locking is an expected outcome of migration theory \citep{CresswellNelson2008,Ogihara2015mig}, the current sample of exoplanets shows only a mild tendency for systems to be
near mean motion commensurabilities \citep{WinnFabrycky2015}. Motivated by
this observational finding, this paper derives an analytic criterion
for turbulent disruption of planetary resonances and demonstrates its
viability through numerical integrations. Our specific results are
outlined below (Section \ref{summary}), followed by a conceptual
interpretation of the calculations (Section \ref{concept}), and finally a
discussion of the implications (Section \ref{discussion}).

\subsection{Summary of Results} 
\label{summary} 

The main result of this paper is the derivation of the constraint
necessary for turbulent fluctuations to compromise mean motion
resonance (given by equation [\ref{criterionchi}]).  This criterion
exhibits a strong dependence on the ratio of planetary mass to stellar
mass, but also has significant dependence on the local surface
density. That is, turbulence can successfully disrupt mean motion
resonances only for systems with sufficiently small mass ratios and/or large
surface densities (see Figure \ref{critfig}).

The analytic estimate (\ref{criterionchi}) for the conditions required
for turbulence to remove planet pairs from resonance has been verified by
numerical integrations. To this end, we have constructed a model of disk-driven resonant dynamics
in the perturbative regime, and have calculated the time evolution of
the resonance promiximity parameter $\varepsilon$ (Section
\ref{perturb}).  The results confirm the analytical prediction that given nominal disk parameters, systems with mass ratios
smaller than $(m_1+m_2)/M\sim10^{-5}\sim 3M_{\oplus}/M_{\odot}$ are forced out of resonance by turbulence, whereas
systems with larger mass ratios survive (Figure \ref{delta}). We have
also performed full $N$-body simulations of the problem (Section
\ref{nbody}). These calculations further indicate that planetary systems with
small mass ratios are readily moved out of resonance by turbulent
fluctuations, whereas systems with larger mass ratios are not (Figure
\ref{Nbodyfig}). Accordingly, the purely analytic treatment, simulations performed within the framework of perturbation
theory, and the full $N$-body experiments all yield consistent results. 

For circumstellar disks with properties comparable to the minimum mass
solar nebula \citep{1981PThPS..70...35H}, the results of this paper suggest that compact \textit{Kepler}-type planetary systems are relatively close to the border-line for stochastic disruption
of primordial mean motion commensurabilities. Nonetheless, with a cumulative mass ratio that typically lies in the range of $(m_1+m_2)/M\sim10^{-5} - 10^{-4}$ (Figure \ref{data}), the majority of these planets are
sufficiently massive that their resonances can survive in the face of
turbulent disruption, provided that the perturbations operate at the expected amplitudes
(this result also assumes that the stochastic fluctuations act over a
time scale that is comparable to the migration time).

Given critical combinations of parameters (for which equation [\ref{criterionchi}] evaluates to a value of order unity), resonant systems can ensue, but
they routinely come out of the disk evolution phase with large libration
amplitudes. This effect has already been pointed out in previous work
\citep{alb,rein,lecoanet,ketchum}, which focused primarily on numerical
simulations with limited analytical characterization. Importantly, this notion suggests that the stochastic forcing mechanism may be critical to setting up
extrasolar planetary systems like GJ\,876 and \textit{Kepler}-36 that exhibit
rapid dynamical chaos \citep{Decketal,BatDeckHol}.

Although this work has mainly focused on the evolution of sub-Jovian planets, we can reasonably speculate that turbulent fluctuations are unlikely to strongly affect mean motion
resonances among giant planets. In addition to having mass ratios well
above the critical limit, the influence that the disk exerts on large
planets is further diminished because of gap-opening \citep{2006Icar..181..587C,2013ApJ...769...41D}.  However, one
complication regarding this issue is that the damping rate of
eccentricity is also reduced due to the gap (e.g., \citealt{DuffellChiang2015}). Since both the excitation
and damping mechanisms are less effective in the gap-opening regime,
a minority of systems could in principle allow for excitation to dominate.  

\subsection{Conceptual Considerations} 
\label{concept} 


The analysis presented herein yields a practical measure that informs the outcome of dynamical evolution of multi-planetary systems embedded in turbulent protoplanetary disks. While numerical experiments confirm that the analytic theory indeed provides an acceptable representation of perturbed $N$-body dynamics, the phenomenological richness inherent to the problem calls for an additional, essentially qualitative account of the results. This is the purpose of the following discussion.

Within the framework of our most realistic description of the relevant physics (i.e., the $N$-body treatment), the effect of turbulent fluctuations is
to provide impulsive changes to the planet velocities. The turbulence
has a coherence time of order one orbital period, so that the
fluctuations provide a new realization of the random gravitational
field on this time scale \citep{alb}. With these impulses, the orbital elements of
the planets, specifically the semi-major axis $a$ and eccentricity
$e$, execute a random walk. In other words, as the elements vary, the changes in $a$ and $e$ accumulate
in a diffusive manner \citep{rein}. Simultaneously, the interactions between planets and the spiral density waves they induce in the nebula lead to smooth changes in the orbital periods, as well as damping of the planetary eccentricities \citep{KleyNelson2012}. 


In contrast with aforementioned disk-driven effects, the bandwidth of a planetary resonance is typically described in terms of maximal libration amplitude of a critical angle $\phi$ that obeys d'Alembert rules (e.g., see Chapter 8 of \citealt{md99}). Thus, the conceptual difficulty lies in connecting how the extrinsic forcing of orbital elements translates to the evolution of this angle. Within the framework of our theoretical model, this link is enabled by the Hamiltonian model of mean motion resonance (equation [\ref{Hammy}]; \citealt{Wisdom1986}).

In the parameter range relevant to the problem at hand, the behavior of Hamiltonian (\ref{Hammy}) is well-approximated by that of a simple pendulum \citep{HenrardLemaitre1983}. Specifically, the equilibrium value of $\varepsilon$ dictates the value of the pendulum's action, $\Phi$, at which zero-amplitude libration of the angle $\phi$ can occur, as well as the location of the separatrix. Correspondingly, oscillation of the angle $\phi$ translates to variations of the action $\Phi$, which is in turn connected to the eccentricities (equations [\ref{AAcoords}]) as well as the semi-major axes, through conservation of the generalized Tisserand parameter \citep{BatMorby2013b}.

In this picture, there are two ways to drive an initially stationary pendulum to circulation: one is to perturb the ball of the pendulum directly (thereby changing the energy-level of the trajectory), and the other is to laterally rock the base (thus modulating the separatrix along the $\Phi$-axis). These processes are directly equivalent to the two types of diffusion considered in our calculations. That is, [1] diffusion in the dynamic variables $x$ and $y$ themselves (explicitly connected to eccentricities and resonant angle) is analogous to direct perturbations to the ball of the pendulum, while [2] diffusion in the proximity parameter $\varepsilon$ (explicitly connected to the semi-major axes) corresponds to shaking the base of the pendulum back and forth.

Meanwhile, consequences of eccentricity damping and convergent migration are equivalent to friction that acts to return the ball of the pendulum back to its undisturbed state, and restore the separatrix to its equilibrium position, respectively. In the type-I migration regime however, eccentricity damping by the disk is far more efficient than orbital decay \citep{Tanaka2004}, meaning that the ball of the pendulum is effectively submerged in water, while the base of the pendulum is only subject to air-resistance (in this analogy). As a result, the latter process --- diffusion in proximity parameter $\varepsilon$ --- ends up being more important for purposes of moving planets out of mean motion resonance (see equation [\ref{avse}]).

\subsection{Discussion}
\label{discussion} 


The work presented herein suggests that turbulent forcing is unlikely to be the single dominant effect that sculpts the final orbital distribution of exoplanets. At the same time, the functional form of expression (\ref{criterionchi}) yields important insight into the evolutionary aspects of the planet formation process. Particularly, because the resonance disruption criterion
depends on the disk mass, it implies a certain time-dependence of the
mechanism itself (as the nebula dissipates, the critical mass ratio below which the mechanism operates
decreases from a value substantially above the Earth-Sun mass
ratio, to one below). This means that even though the turbulent disruption mechanism becomes ineffective
in a weaning nebula, it may be key to facilitating growth in the early stages of evolution of
planetary systems, by allowing pairs of proto-planets to skip over mean-motion commensurabilities and merge, instead of forming resonant chains. In essence, this type of dynamical behavior is seen in the large-scale numerical experiments of \citet{lyra}.

For much of this work, the system parameters that we use effectively
assume a maximum rate of orbital convergence. Because the quantitative
nature of migration can change substantially in the inner nebula, the
actual rate of orbital convergence may be somewhat lower
\citep{Paardekooper,2015A&A...575A..28B}.  This change would make planetary resonances more
susceptible to stochastic disruption. At the same time, we have not
taken into account the inhibition of the random gravitational field
through non-ideal magnetohydrodynamic effects \citep{ormelpaper},
which would weaken the degree of stochastic forcing. Both of these
effects can be incorporated into the criterion of equation
(\ref{criterionchi}) by lowering the migration factor $f$ and the
value of $\alpha$ accordingly.  However, because both of these quantities
appear under a square root in the expression, the sensitivity
of our results to these corrections is not expected to be extreme.

This work assumes that turbulence operates in circumstellar
disks at the expected levels. The presence of turbulence is most commonly attributed to the magneto-rotational-instability \citep{mri},
which in turn requires the disk to be sufficiently ionized. Although
the innermost regions of the disk are expected to be ionized by
thermal processes, dead zones could exist in intermediate part of the
disk (\citealt{gammie}; see also \citealt{2013ApJ...769...76B}), and ionization by cosmic rays can be suppressed
in the outer disk \citep{cleevesone}. Indeed, suppressed levels of
ionization are now inferred from ALMA observations of young star/disk
systems \citep{cleeves}, implying that the assumption of sufficient
ionization --- and hence active MRI turbulence --- is not guarenteed. At the same time, our model is agnostic towards the origins of turbulent fluctuations themselves, and can be employed equally well if a purely hydrodynamic source of turbulence were responsible for angular momentum transport within the nebula \citep{2013MNRAS.435.2610N,2015ApJ...811...17L}.

In light of the aforementioned uncertainties inherent to the problem at hand, it is of considerable interest to explore if simply adjusting the parameters can, in principle, yield consistency between the model and the observations. That is, can reasonable changes to the migration rate, etc., generate agreement between the turbulent resonance disruption hypothesis and the data? Using equation (\ref{criterionchi}), we find that increasing the local surface density by an order of magnitude ($\Sigma=10\langle\Sigma\rangle=170,000\,$g/cm$^2$) while lowering the orbital convergence rate a hundred-fold ($f=0.01$) and retaining $h/r=0.05$, $\langle a\rangle=0.1\,$AU, $\alpha=0.01$ yields $(m_1+m_2)/M\simeq2\times 10^{-4}\sim60M_{\oplus}/M_{\odot}$ as the critical mass ratio, thus explaining the full range of values shown in Figure \ref{data}. Correspondingly, rough agreement between observations and the stochastic migration scenario is reproduced in the work of \citet{Rein2012}, where the amplitude of turbulent forcing was tuned to give consistency with data.

Although this line of reasoning may appear promising, it is important to note that as the disk accretes onto the star, the local surface density will diminish, causing the critical mass ratio to decrease as well. Meanwhile, even with a reduction factor of $f=0.01$, the type-I migration timescale remains shorter than the $\sim$few Myr lifetime of the nebula, as long as $\Sigma\gtrsim0.1\langle\Sigma\rangle=170\,$g/cm$^2$. As a result, we argue that any realistic distribution of the assumed parameters is unlikely to allow turbulence to provide enough resonance disruption to explain the entire set of observations.

If disk turbulence does not play the defining role in generating an
observational census of extrasolar planets that is neither dominated
by, nor devoid of, mean motion resonances, than what additional
processes are responsible for the extant data set? As already
mentioned in the introduction, there are two other ways in which
planets can avoid resonant locking -- resonant metastability \citep{GoldShicht2014,2016arXiv161106463X} and capture
probability suppression \citep{Batygin2015}.  The first mechanism requires that the outer
planet is more massive then the inner planet to compromise
resonance \citep{DeckBatygin}. As a result, observed resonant systems would almost always
have a more massive inner planet, but this ordering is not reflected
in the data. On the other hand, the (second) capture suppression
mechanism requires disk eccentricities of order
$e_{\rm{disk}}\sim0.02$ to explain the data. Importantly, disk eccentricities of
this magnitude (and greater) are not only an expected result of
theoretical calculations, they are invoked to explain observations of
asymmetric glow of dust \citep{2015ApJ...798L..25M}.

In conclusion, turbulent fluctuations probably do not explain the
entire ensemble of observed planetary systems, which exhibit only a weak preference for mean motion commensurability. In addition to
turbulent forcing, many other physical processes are likely at work,
where perhaps the most promising mechanism is capture suppression due to nonzero
disk eccentricities. Nonetheless, a subset of exotic planetary systems
that exhibit large-amplitude resonant librations likely require a
turbulent origin. The relative duty cycle of this mechanism, and
others, poses an interesting problem for further exploration.

\medskip 

\textbf{Acknowledgments:} We would like to thank Juliette Becker, Tony Bloch, Wlad Lyra and Chris Spalding for useful discussions, as well as the referee, Hanno Rein, whose insightful report led to a considerable improvement of the manuscript. K.B. acknowledges support from the NSF AAG program AST1517936, and from Caltech. F.C.A. acknowledges support from the NASA Exoplanets Research Program NNX16AB47G, and from the University of Michigan.


\newpage 

\begin{thebibliography} 

\bibitem[Adams et al.(2008)]{alb}
Adams, F. C., Laughlin, G., \& Bloch, A. M. 2008, ApJ, 683, 1117

\bibitem[Allan(1969)]{Allan1969} 
Allan, R. R. 1969, AJ, 74, 497

\bibitem[Allan(1970)]{Allan1970}
Allan, R. R. 1970, Celest. Mech., 2, 121

\bibitem[Balbus \& Hawley(1991)]{mri} 
Balbus, S. A., \& Hawley, J. F. 1991, ApJ, 376, 214

\bibitem[Bai \& Stone(2013)]{2013ApJ...769...76B} Bai, X.-N., \& Stone, J.~M.\ 2013, \apj, 769, 76 

\bibitem[Batalha et al.(2013)]{Batalha} 
Batalha, N. M. et al. 2013, ApJS, 204, 24

\bibitem[Batygin(2015)]{Batygin2015} 
Batygin, K. 2015, MNRAS, 451, 2589

\bibitem[Batygin et al.(2015)]{BatDeckHol} 
Batygin, K., Deck, K. M., \& Holman, M. J. 2015, AJ, 149, 167

\bibitem[Batygin et al. (2011)]{batynmorbid}
Batygin, K., \& Morbidelli, A. 2011, {Celest. Mech.}, 111, 219

\bibitem[Batygin \& Morbidelli(2013)]{BatMorby2013b}
Batygin, K., \& Morbidelli, A. 2013, A\&A, 556, A28

\bibitem[Bitsch \& Kley(2011)]{Bitsch}
Bitsch, B., \& Kley, W. 2011, A\&A, 536, A77

\bibitem[Bitsch et al.(2015)]{2015A&A...575A..28B} Bitsch, B., Johansen, A., Lambrechts, M., \& Morbidelli, A.\ 2015, \aap, 575, A28 

\bibitem[Borders \& Goldreich(1984)]{BordersGoldreich1984}
Borderies, N., \& Goldreich, P. 1984, Celest. Mech., 32, 127

\bibitem[Chiang \& Laughlin(2013)]{ChiangLaughlin}
Chiang, E., \& Laughlin, G. 2013, MNRAS, 431, 3444 

\bibitem[Cleeves et al.(2013)]{cleevesone} 
Cleeves, L. I., Adams, F. C., \& Bergin, E. A. 2013, ApJ, 772, 5 

\bibitem[Cleeves et al.(2015)]{cleeves} 
Cleeves, L. I., Bergin, E. A., Qi, C., Adams, F. C., \& {\"O}berg, K. I.
2015, ApJ, 799, 204 

\bibitem[Coleman \& Nelson(2016)]{ColemanNelson}
Coleman, G.A.L., \& Nelson, R. P. 2016, MNRAS, 457, 2480

\bibitem[Cresswell \& Nelson(2008)]{CresswellNelson2008} 
Cresswell, P., \& Nelson, R. P. 2008, A\&A, 482, 677

\bibitem[Crida et al.(2006)]{2006Icar..181..587C} Crida, A., Morbidelli, A., \& Masset, F.\ 2006, Icarus, 181, 587 

\bibitem[Crida et al.(2008)]{Crida2008}
Crida, A., S{\'a}ndor, A., \& Kley, W., 2008, A\&A, 483, 325 

\bibitem[D'Angelo \& Bodenheimer(2016)]{DAngeloBodenheimer2016} D'Angelo, G., \& Bodenheimer, P.\ 2016, \apj, 828, 33 

\bibitem[Deck et al.(2012)]{Decketal} 
Deck, K. M. et al. 2012, ApJ, 755, 21  

\bibitem[Deck et al.(2013)]{Deck2013}
Deck, K. M., Payne, M., \& Holman, M. J. 2013, ApJ, 774, 129 

\bibitem[Deck \& Batygin(2015)]{DeckBatygin} 
Deck, K. M., \& Batygin, K. 2015, ApJ, 810, 119 

\bibitem[Duffell \& MacFadyen(2013)]{2013ApJ...769...41D} Duffell, P.~C., \& MacFadyen, A.~I.\ 2013, \apj, 769, 41 


\bibitem[Duffell \& Chiang(2015)]{DuffellChiang2015} Duffell, P.~C., \& Chiang, E.\ 2015, \apj, 812, 94 


\bibitem[Fabrycky et al.(2014)]{Fabrycky2014}
Fabrycky, D. C. et al. 2014, ApJ, 790, 146

\bibitem[Foreman-Mackey et al.(2014)]{Forman-Mackey} 
Foreman-Mackey, D., Hogg, D. W., \& Morton, T. D. 2014, ApJ,
795, 64 

\bibitem[Fressin et al.(2013)]{Fressin2013} Fressin, F., Torres, G., Charbonneau, D., et al.\ 2013, \apj, 766, 81 

\bibitem[Gammie(1996)]{gammie} 
Gammie, C. F. 1996, ApJ, 457, 355

\bibitem[Goldreich(1965)]{Goldreich1965} 
Goldreich, P. 1965, MNRAS, 130, 159

\bibitem[Goldreich \& Schlichting(2014)]{GoldShicht2014} 
Goldreich, P., \& Schlichting, H. E. 2014, AJ, 147, 32

\bibitem[Goldreich \& Tremaine(1980)]{GoldreichTremaine1980} 
Goldreich, P., \& Tremaine, S. 1980, ApJ, 241, 425

\bibitem[Goldreich \& Tremaine(1982)]{GoldreichTremaineRing}  
Goldreich, P., \& Tremaine, S. 1982, ARA\&A, 20, 249  

\bibitem[Hansen \& Murray(2015)]{MurrayHansen}
Hansen, B.M.S., \& Murray, N. 2015, MNRAS, 448, 1044 

\bibitem[Hayashi(1981)]{1981PThPS..70...35H} Hayashi, C.\ 1981, Progress of Theoretical Physics Supplement, 70, 35 

\bibitem[Henrard(1986)]{Henrard1986} 
Henrard, J., \& Lemaitre, A. 1986, Celest. Mech., 39, 213 

\bibitem[Henrard \& Lamaitre(1983)]{HenrardLemaitre1983} Henrard, J., \& Lamaitre, A.\ 1983, Celestial Mechanics, 30, 197 

\bibitem[Horn et al.(2012)]{lyra} 
Horn, B., Lyra, W., Mac Low, M-M., \& S{\'a}ndor, Z. 2012, ApJ, 750, 34  

\bibitem[Howard et al.(2012)]{Howard}
Howard, A. W. et al. 2012, ApJS, 201, 15

\bibitem[Johnson et al.(2006)]{johnson} 
Johnson, E. T., Goodman, J., \& Menou, K. 2006, ApJ, 647, 1413

\bibitem[Jontof-Hutter et al.(2014)]{2014ApJ...785...15J} Jontof-Hutter, D., Lissauer, J.~J., Rowe, J.~F., \& Fabrycky, D.~C.\ 2014, \apj, 785, 15 


\bibitem[Ketchum et al.(2011)]{ketchum} 
Ketchum, J. A., Adams, F. C., \& Bloch, A. M. 2011, ApJ, 726, 53   

\bibitem[Kley \& Nelson(2012)]{KleyNelson2012}
Kley, W., \& Nelson, R. P. 2012, ARA\&A, 50, 211

\bibitem[Laughlin et al.(2004)]{lsa}
Laughlin, G., Steinacker, A., \& Adams, F. C. 2004, ApJ, 608, 489

\bibitem[Lecoanet et al.(2009)]{lecoanet}
Lecoanet, D., Adams, F. C., \& Bloch, A. M. 2009, ApJ, 692, 659  

\bibitem[Lee \& Peale(2002)]{leepeale}
Lee, M.-H., \& Peale, S. J. 2002, ApJ, 567, 596

\bibitem[Lee \& Chiang(2015)]{Lee2015} 
Lee, E. J., \& Chiang, E. 2015, ApJ, 811, 41 

\bibitem[Lee \& Chiang(2016)]{Lee2016} 
Lee, E. J., \& Chiang, E. 2016, ApJ, 817, 90

\bibitem[Lin \& Youdin(2015)]{2015ApJ...811...17L} Lin, M.-K., \& Youdin, A.~N.\ 2015, \apj, 811, 17 

\bibitem[Malhotra(1993)]{Malhotra1993}
Malhotra, R. 1993, Nature, 365, 819

\bibitem[Mestel(1963)]{Mestel63}
Mestel, L. 1963, MNRAS, 126, 553

\bibitem[Mills et al.(2016)]{2016Natur.533..509M} Mills, S.~M., Fabrycky, D.~C., Migaszewski, C., et al.\ 2016, \nat, 533, 509 

\bibitem[Mittal \& Chiang(2015)]{2015ApJ...798L..25M} Mittal, T., \& Chiang, E.\ 2015, \apjl, 798, L25 

\bibitem[Morbidelli(2002)]{Morby} 
Morbidelli, A. 2002, Modern Celestial Mechanics: Aspects of Solar 
System Dynamics (London: Taylor \& Francis) 

\bibitem[Mulders et al.(2015)]{Mulders2015} Mulders, G.~D., Pascucci, I., \& Apai, D.\ 2015, \apj, 798, 112

\bibitem[Murray \& Dermott(1999)]{md99}
Murray, C. D., \& Dermott, S. F. 1999, {Solar System Dynamics} 
(Cambridge: Cambridge Univ. Press) 

\bibitem[Nelson \& Papaloizou(2004)]{nelson}
Nelson R. P., \& Papaloizou J.C.B. 2004, MNRAS, 350, 849

\bibitem[Nelson et al.(2013)]{2013MNRAS.435.2610N} Nelson, R.~P., Gressel, O., \& Umurhan, O.~M.\ 2013, \mnras, 435, 2610 

\bibitem[Nesvorn{\'y} \& Morbidelli(2008)]{morbyone}  
Nesvorn{\'y}, D., \& Morbidelli, A. 2008, Icarus, 688, 636 

\bibitem[Nobili \& Roxburgh(1986)]{genrel} Nobili, A., \& Roxburgh, I. W. 1986, IAUS, 114, 105

\bibitem[Ogihara et al.(2015)]{Ogihara2015insitu} Ogihara, M., Morbidelli, A., \& Guillot, T.\ 2015, \aap, 578, A36 

\bibitem[Ogihara et al.(2015)]{Ogihara2015mig} Ogihara, M., Morbidelli, A., \& Guillot, T.\ 2015, \aap, 584, L1 


\bibitem[Okuzumi \& Hirose(2011)]{okuzumi} 
Okuzumi, S., \& Hirose, S. 2011, ApJ, 742, 65 

\bibitem[Okuzumi \& Ormel(2013)]{OkuzumiOrmel2013} Okuzumi, S., \& Ormel, C.~W.\ 2013, \apj, 771, 43 


\bibitem[{\"O}pik(1976)]{Opik1976}
Opik, E. J. 1976, Interplanetary Encounters: 
Close-range gravitational interactions (New York: Elsevier) 

\bibitem[Ormel \& Okuzumi(2013)]{ormelpaper}
Ormel, C. W., \& Okuzumi, S. 2013, ApJ, 771, 44

\bibitem[Paardekooper(2014)]{Paardekooper} Paardekooper, S.-J.\ 2014, \mnras, 444, 2031 

\bibitem[Papaloizou \& Larwood(2000)]{PapaloizouLarwood2000}
Papaloizou, J.C.B., \& Larwood, J. D. 2000, MNRAS, 315, 823

\bibitem[Peale(1976)]{Peale1976}
Peale, S. J. 1976, ARA\&A, 14, 215

\bibitem[Petigura et al.(2013)]{Petigura} 
Perigura, E. A., Howard, A. W., \& Marcy, G. W. 2013, PNAS, 110, 19273

\bibitem[Press et al.(1992)]{Press1992} 
Press, W. H., Teukolsky, S. A., Vetterling, W. T., \& Flannery,
B. P. 1992, Numerical Recipes in FORTRAN: The Art of Scientific
Computing (Cambridge: Cambridge Univ. Press)

\bibitem[Quillen(2006)]{quill}
Quillen, A. C. 2006, MNRAS, 365, 1367 

\bibitem[Rein \& Papaloizou(2009)]{rein} 
Rein, H., \& Papaloizou, J.C.P. 2009, A\&A, 497, 595  

\bibitem[Rein et al.(2010)]{2010A&A...510A...4R} Rein, H., Papaloizou, J.~C.~B., \& Kley, W.\ 2010, \aap, 510, A4 

\bibitem[Rein(2012)]{Rein2012} Rein, H.\ 2012, \mnras, 427, L21 

\bibitem[Rivera et al.(2010)]{2010ApJ...719..890R} Rivera, E.~J., Laughlin, G., Butler, R.~P., et al.\ 2010, \apj, 719, 890 


\bibitem[Rogers(2015)]{Rogers2015} 
Rogers, L. A. 2015, ApJ, 801, 41 

\bibitem[Schlichting(2014)]{Schlichting}
Schlighting, H. 2014, ApJ, 795, 15 

\bibitem[Sessin \& Ferraz-Mello(1984)]{SessinFerraz-Mello1984}
Sessin, W., \& Ferraz-Mello, S. 1984, Celest. Mech., 32, 307

\bibitem[Sinclair(1970)]{Sinclair1970}
Sinclair, A. T. 1970, MNRAS, 148, 325

\bibitem[Sinclair(1972)]{Sinclair1972}
Sinclair, A. T. 1972, MNRAS, 160, 169

\bibitem[Shakura \& Sunyaev(1973)]{ShakuraSunayev1973}
Shakura, N. I., \& Sunyaev, R. A. 1973, A\&A, 24, 337
 
\bibitem[Tanaka et al.(2002)]{Tanaka2002}
Tanaka, H., Takeuchi, T., Ward, W. R. 2002, ApJ, 565, 1257

\bibitem[Tanaka \& Ward(2004)]{Tanaka2004} 
Tanaka, H., \& Ward, W. R. 2004, ApJ, 602, 388 

\bibitem[Terquem \& Papaloizou(2007)]{TerquemPap2007}
Terquem, C., Papaloizou, J.C.B. 2007, ApJ, 654, 1110

\bibitem[Van Kampen(2001)]{vankampen}
Van Kampen, N. G. 2001, Stochastic Processes in Physics 
and Chemistry (Amsterdam: North Holland) 

\bibitem[Weiss \& Marcy(2014)]{WeissMarcy2014} 
Weiss, L. M., \& Marcy, G. W. 2014, ApJ, 783, 6 

\bibitem[Winn \& Fabrycky(2015)]{WinnFabrycky2015} Winn, J.~N., \& Fabrycky, D.~C.\ 2015, \araa, 53, 409 

\bibitem[Xu \& Lai(2016)]{2016arXiv161106463X} Xu, W., \& Lai, D.\ 2016, arXiv:1611.06463 

\bibitem[Wisdom(1986)]{Wisdom1986}
Wisdom, J. 1986, Celest. Mech., 38, 175

\end{thebibliography}
\end{document}